

%
%


\def\famname{
 \textfont0=\textrm \scriptfont0=\scriptrm
 \scriptscriptfont0=\sscriptrm
 \textfont1=\textmi \scriptfont1=\scriptmi
 \scriptscriptfont1=\sscriptmi
 \textfont2=\textsy \scriptfont2=\scriptsy \scriptscriptfont2=\sscriptsy
 \textfont3=\textex \scriptfont3=\textex \scriptscriptfont3=\textex
 \textfont4=\textbf \scriptfont4=\scriptbf \scriptscriptfont4=\sscriptbf
 \skewchar\textmi='177 \skewchar\scriptmi='177
 \skewchar\sscriptmi='177
 \skewchar\textsy='60 \skewchar\scriptsy='60
 \skewchar\sscriptsy='60
 \def\rm{\fam0 \textrm} \def\bf{\fam4 \textbf}}
\def\sca#1{scaled\magstep#1} \def\scah{scaled\magstephalf} 
\def\twelvepoint{
 \font\textrm=cmr12 \font\scriptrm=cmr8 \font\sscriptrm=cmr6
 \font\textmi=cmmi12 \font\scriptmi=cmmi8 \font\sscriptmi=cmmi6 
 \font\textsy=cmsy10 \sca1 \font\scriptsy=cmsy8
 \font\sscriptsy=cmsy6
 \font\textex=cmex10 \sca1
 \font\textbf=cmbx12 \font\scriptbf=cmbx8 \font\sscriptbf=cmbx6
 \font\it=cmti12
 \font\sectfont=cmbx12 \sca1
 \font\sectmath=cmmib10 \sca2
 \font\sectsymb=cmbsy10 \sca2
 \font\refrm=cmr10 \scah \font\refit=cmti10 \scah
 \font\refbf=cmbx10 \scah
 \def\twelverm{\textrm} \def\twelveit{\it} \def\twelvebf{\textbf}
 \famname \textrm 
 \advance\voffset by .06in \advance\hoffset by .28in
 \normalbaselineskip=17.5pt plus 1pt \baselineskip=\normalbaselineskip
 \parindent=21pt
 \setbox\strutbox=\hbox{\vrule height10.5pt depth4pt width0pt}}


\catcode`@=11

{\catcode`\'=\active \def'{{}^\bgroup\prim@s}}

\def\screwcount{\alloc@0\count\countdef\insc@unt}   
\def\screwdimen{\alloc@1\dimen\dimendef\insc@unt} 
\def\screwbox{\alloc@4\box\chardef\insc@unt}

\catcode`@=12


\overfullrule=0pt			
\vsize=9in \hsize=6in
\lineskip=0pt				
\abovedisplayskip=1.2em plus.3em minus.9em 
\belowdisplayskip=1.2em plus.3em minus.9em	
\abovedisplayshortskip=0em plus.3em	
\belowdisplayshortskip=.7em plus.3em minus.4em	
\parindent=21pt
\setbox\strutbox=\hbox{\vrule height10.5pt depth4pt width0pt}
\def\makefootline{\baselineskip=30pt \line{\the\footline}}
\footline={\ifnum\count0=1 \hfil \else\hss\twelverm\folio\hss \fi}
\pageno=1


\def\put(#1,#2)#3{\screwdimen\unit  \unit=1in
	\vbox to0pt{\kern-#2\unit\hbox{\kern#1\unit
	\vbox{#3}}\vss}\nointerlineskip}


\def\\{\hfil\break}
\def\newpage{\vfill\eject}
\def\center{\leftskip=0pt plus 1fill \rightskip=\leftskip \parindent=0pt
 \def\textindent##1{\par\hangindent21pt\footrm\noindent\hskip21pt
 \llap{##1\enspace}\ignorespaces}\par}
\def\unnarrower{\leftskip=0pt \rightskip=\leftskip}


\def\vol#1 {{\refbf#1} }		 


\def\NP #1 {{\refit Nucl. Phys.} {\refbf B{#1}} }
\def\PL #1 {{\refit Phys. Lett.} {\refbf{#1}} }
\def\PR #1 {{\refit Phys. Rev. Lett.} {\refbf{#1}} }
\def\PRD #1 {{\refit Phys. Rev.} {\refbf D{#1}} }


\hyphenation{pre-print}
\hyphenation{quan-ti-za-tion}

%
%


\def\oonoo#1#2#3{\vbox{\ialign{##\crcr
	\hfil\hfil\hfil{$#3{#1}$}\hfil\crcr\noalign{\kern1pt\nointerlineskip}
	$#3{#2}$\crcr}}}
\def\oon#1#2{\mathchoice{\oonoo{#1}{#2}{\displaystyle}}
	{\oonoo{#1}{#2}{\textstyle}}{\oonoo{#1}{#2}{\scriptstyle}}
	{\oonoo{#1}{#2}{\scriptscriptstyle}}}
\def\dt#1{\oon{\hbox{\bf .}}{#1}}  
\def\ddt#1{\oon{\hbox{\bf .\kern-1pt.}}#1}    
\def\slap#1#2{\setbox0=\hbox{$#1{#2}$}
	#2\kern-\wd0{\hfuzz=1pt\hbox to\wd0{\hfil$#1{/}$\hfil}}}
\def\sla#1{\mathpalette\slap{#1}}                
\def\bop#1{\setbox0=\hbox{$#1M$}\mkern1.5mu
	\lower.02\ht0\vbox{\hrule height0pt depth.06\ht0
	\hbox{\vrule width.06\ht0 height.9\ht0 \kern.9\ht0
	\vrule width.06\ht0}\hrule height.06\ht0}\mkern1.5mu}
\def\bo{{\mathpalette\bop{}}}                        
\def~{\widetilde} 
\mathcode`\*="702A                  
\def\in{\relax\ifmmode\mathchar"3232\else{\refit in\/}\fi} 
\def\f#1#2{{\textstyle{#1\over#2}}}	   
\def\half{{\textstyle{1\over{\raise.1ex\hbox{$\scriptstyle{2}$}}}}}

\def\Gamma{\mathchar"0100}
\def\Delta{\mathchar"0101}
\def\Theta{\mathchar"0102}
\def\Lambda{\mathchar"0103}
\def\Xi{\mathchar"0104}
\def\Pi{\mathchar"0105}
\def\Sigma{\mathchar"0106}
\def\Upsilon{\mathchar"0107}
\def\Phi{\mathchar"0108}
\def\Psi{\mathchar"0109}
\def\Omega{\mathchar"010A}

\catcode128=13 \def €{\"A}                 
\catcode129=13 \def {\AA}                 
\catcode130=13 \def '{\c}           	   
\catcode131=13 \def ƒ{\'E}                   
\catcode132=13 \def "{\~N}                   
\catcode133=13 \def …{\"O}                 
\catcode134=13 \def †{\"U}                  
\catcode135=13 \def ‡{\'a}                  
\catcode136=13 \def ˆ{\`a}                   
\catcode137=13 \def ‰{\^a}                 
\catcode138=13 \def Š{\"a}                 
\catcode139=13 \def ‹{\~a}                   
\catcode140=13 \def Œ{\alpha}            
\catcode141=13 \def {\chi}                
\catcode142=13 \def Ž{\'e}                   
\catcode143=13 \def {\`e}                    
\catcode144=13 \def {\^e}                  
\catcode145=13 \def '{\"e}                
\catcode146=13 \def '{\'\i}                 
\catcode147=13 \def "{\`\i}                  
\catcode148=13 \def "{\^\i}                
\catcode149=13 \def •{\"\i}                
\catcode150=13 \def –{\~n}                  
\catcode151=13 \def —{\'o}                 
\catcode152=13 \def ˜{\`o}                  
\catcode153=13 \def ™{\^o}                
\catcode154=13 \def š{\"o}                 
\catcode155=13 \def ›{\~o}                  
\catcode156=13 \def œ{\'u}                  
\catcode157=13 \def {\`u}                  
\catcode158=13 \def ž{\^u}                
\catcode159=13 \def Ÿ{\"u}                
\catcode160=13 \def  {\tau}               
\catcode161=13 \mathchardef ¡="2203     
\catcode162=13 \def ¢{\oplus}           
\catcode163=13 \def £{\relax\ifmmode\to\else\itemize\fi} 
\catcode164=13 \def ¤{\subset}	  
\catcode165=13 \def ¥{\infty}           
\catcode166=13 \def ¦{\mp}                
\catcode167=13 \def §{\sigma}           
\catcode168=13 \def ¨{\rho}               
\catcode169=13 \def ©{\gamma}         
\catcode170=13 \def ª{\leftrightarrow} 
\catcode171=13 \def «{\relax\ifmmode\acute\else\expandafter\'\fi}
\catcode172=13 \def ¬{\relax\ifmmode\expandafter\ddt\else\expandafter\"\fi}
\catcode173=13 \def ­{\equiv}            
\catcode174=13 \def ®{\approx}          
\catcode175=13 \def ¯{\Omega}          
\catcode176=13 \def °{\otimes}          
\catcode177=13 \def ±{\ne}                 
\catcode178=13 \def ²{\le}                   
\catcode179=13 \def ³{\ge}                  
\catcode180=13 \def ´{\upsilon}          
\catcode181=13 \def µ{\mu}                
\catcode182=13 \def ¶{\delta}             
\catcode183=13 \def ·{\epsilon}          
\catcode184=13 \def ¸{\Pi}                  
\catcode185=13 \def ¹{\pi}                  
\catcode186=13 \def º{\beta}               
\catcode187=13 \def »{\partial}           
\catcode188=13 \def ¼{\nobreak\ }       
\catcode189=13 \def ½{\zeta}               
\catcode190=13 \def ¾{\sim}                 
\catcode191=13 \def ¿{\omega}           
\catcode192=13 \def À{\dt}                     
\catcode193=13 \def Á{\gets}                
\catcode194=13 \def Â{\lambda}           
\catcode195=13 \def Ã{\nu}                   
\catcode196=13 \def Ä{\phi}                  
\catcode197=13 \def Å{\xi}                     
\catcode198=13 \def Æ{\psi}                  
\catcode199=13 \def Ç{\int}                    
\catcode200=13 \def È{\oint}                 
\catcode201=13 \def É{\relax\ifmmode\cdot\else\vol\fi}    
\catcode202=13 \def Ê{\relax\ifmmode\,\else\thinspace\fi}
\catcode203=13 \def Ë{\`A}                      
\catcode204=13 \def Ì{\~A}                      
\catcode205=13 \def Í{\~O}                      
\catcode206=13 \def Î{\Theta}              
\catcode207=13 \def Ï{\theta}               
\catcode208=13 \def Ð{\relax\ifmmode\bar\else\expandafter\=\fi}
\catcode209=13 \def Ñ{\overline}             
\catcode210=13 \def Ò{\langle}               
\catcode211=13 \def Ó{\relax\ifmmode\{\else\ital\fi}      
\catcode212=13 \def Ô{\rangle}               
\catcode213=13 \def Õ{\}}                        
\catcode214=13 \def Ö{\sla}                      
\catcode215=13 \def ×{\relax\ifmmode\check\else\expandafter\v\fi}
\catcode216=13 \def Ø{\"y}                     
\catcode217=13 \def Ù{\"Y}  		    
\catcode218=13 \def Ú{\Leftarrow}       
\catcode219=13 \def Û{\Leftrightarrow}       
\catcode220=13 \def Ü{\relax\ifmmode\Rightarrow\else\sect\fi}
\catcode221=13 \def Ý{\sum}                  
\catcode222=13 \def Þ{\prod}                 
\catcode223=13 \def ß{\widehat}              
\catcode224=13 \def à{\pm}                     
\catcode225=13 \def á{\nabla}                
\catcode226=13 \def â{\quad}                 
\catcode227=13 \def ã{\in}               	
\catcode228=13 \def ä{\star}      	      
\catcode229=13 \def å{\sqrt}                   
\catcode230=13 \def æ{\^E}			
\catcode231=13 \def ç{\Upsilon}              
\catcode232=13 \def è{\"E}    	   	 
\catcode233=13 \def é{\`E}               	  
\catcode234=13 \def ê{\Sigma}                
\catcode235=13 \def ë{\Delta}                 
\catcode236=13 \def ì{\Phi}                     
\catcode237=13 \def í{\`I}        		   
\catcode238=13 \def î{\iota}        	     
\catcode239=13 \def ï{\Psi}                     
\catcode240=13 \def ð{\times}                  
\catcode241=13 \def ñ{\Lambda}             
\catcode242=13 \def ò{\cdots}                
\catcode243=13 \def ó{\^U}			
\catcode244=13 \def ô{\`U}    	              
\catcode245=13 \def õ{\bo}                       
\catcode246=13 \def ö{\relax\ifmmode\hat\else\expandafter\^\fi}
\catcode247=13 \def÷{\relax\ifmmode\tilde\else\expandafter\~\fi}
\catcode248=13 \def ø{\ll}                         
\catcode249=13 \def ù{\gg}                       
\catcode250=13 \def ú{\eta}                      
\catcode251=13 \def û{\kappa}                  
\catcode252=13 \def ü{\half}     		 
\catcode253=13 \def ý{\Gamma} 		
\catcode254=13 \def þ{\Xi}   			
\catcode255=13 \def ÿ{\relax\ifmmode{}^{\dagger}{}\else\dag\fi}


\def\ital#1Õ{{\it#1\/}}	     
\def\un#1{\relax\ifmmode\underline#1\else $\underline{\hbox{#1}}$
	\relax\fi}

\def\roonoo#1#2#3{\vbox{\ialign{##\crcr
	\hfil{$#3{#1}$}\hfil\crcr\noalign{\kern1pt\nointerlineskip}
	$#3{#2}$\crcr}}}

\def\tdt#1{\oon{\hbox{\bf .\kern-1pt.\kern-1pt.}}#1}   
\def\({\eqno(}
\def\li{\openup1\jot \eqalignno}


\def\õ#1{
	\screwcount\num
	\num=1
	\screwdimen\downsy
	\downsy=-1.5ex
	\mkern-3.5mu
	õ
	\loop
	\ifnum\num<#1
	\llap{\raise\num\downsy\hbox{$õ$}}
	\advance\num by1
	\repeat}
\def\upõ#1#2{\screwcount\numup
	\numup=#1
	\advance\numup by-1
	\screwdimen\upsy
	\upsy=.75ex
	\mkern3.5mu
	\raise\numup\upsy\hbox{$#2$}}



\newcount\marknumber	\marknumber=1
\newcount\countdp \newcount\countwd \newcount\countht 

%
%
\ifx\pdfoutput\undefined
\def\rgboo#1{}
\input epsf

\def\postscript#1{\special{" #1}}		
\postscript{
	/bd {bind def} bind def
	/fsd {findfont exch scalefont def} bd
	/sms {setfont moveto show} bd
	/ms {moveto show} bd
	/pdfmark where		
	{pop} {userdict /pdfmark /cleartomark load put} ifelse
	[ /PageMode /UseOutlines		
	/DOCVIEW pdfmark}
\def\bookmark#1#2{\postscript{		
	[ /Dest /MyDest\the\marknumber /View [ /XYZ null null null ] /DEST pdfmark
	[ /Title (#2) /Count #1 /Dest /MyDest\the\marknumber /OUT pdfmark}%
	\advance\marknumber by1}
\def\pdfklink#1#2{%
	\hskip-.25em\setbox0=\hbox{#1}%
		\countdp=\dp0 \countwd=\wd0 \countht=\ht0%
		\divide\countdp by65536 \divide\countwd by65536%
			\divide\countht by65536%
		\advance\countdp by1 \advance\countwd by1%
			\advance\countht by1%
		\def\linkdp{\the\countdp} \def\linkwd{\the\countwd}%
			\def\linkht{\the\countht}%
	\postscript{
		[ /Rect [ -1.5 -\linkdp.0 0\linkwd.0 0\linkht.5 ] 
		/Border [ 0 0 0 ]
		/Action << /Subtype /URI /URI (#2) >>
		/Subtype /Link
		/ANN pdfmark}{\rgb{1 0 0}{#1}}}
%
%
\else
\def\rgboo#1{\pdfliteral{#1 rg #1 RG}}

\pdfcatalog{/PageMode /UseOutlines}		
\def\bookmark#1#2{
	\pdfdest num \marknumber xyz
	\pdfoutline goto num \marknumber count #1 {#2}
	\advance\marknumber by1}
\def\pdfklink#1#2{%
	\noindent\pdfstartlink user
		{/Subtype /Link
		/Border [ 0 0 0 ]
		/A << /S /URI /URI (#2) >>}{\rgb{1 0 0}{#1}}%
	\pdfendlink}
\fi

\def\rgbo#1#2{\rgboo{#1}#2\rgboo{0 0 0}}
\def\rgb#1#2{\mark{#1}\rgbo{#1}{#2}\mark{0 0 0}}
\def\pdflink#1{\pdfklink{#1}{#1}}
\def\xxxlink#1{\pdfklink{[arXiv:#1]}{http://arXiv.org/abs/#1}}

\catcode`@=11

\def\wlog#1{}	


\def\makeheadline{\vbox to\z@{\vskip-36.5\p@
	\line{\vbox to8.5\p@{}\the\headline%
	\ifnum\pageno=\z@\rgboo{0 0 0}\else\rgboo{\topmark}\fi%
	}\vss}\nointerlineskip}
\headline={
	\ifnum\pageno=\z@
		\hfil
	\else
		\ifnum\pageno<\z@
			\ifodd\pageno
				\tenrm\romannumeral-\pageno\hfil\lefthead\hfil
			\else
				\tenrm\hfil\righthead\hfil\romannumeral-\pageno
			\fi
		\else
			\ifodd\pageno
				\tenrm\hfil\righthead\hfil\number\pageno
			\else
				\tenrm\number\pageno\hfil\lefthead\hfil
			\fi
		\fi
	\fi}

\catcode`@=12

\def\righthead{\hfil} \def\lefthead{\hfil}
\nopagenumbers


\def\chrulefill{\rgb{1 0 0}{\hrulefill}}
\def\cdotfill{\rgb{1 0 0}{\dotfill}}
\newcount\area	\area=1
\newcount\cross	\cross=1
\def\volume#1\par{\newpage\noindent{\biggest{\rgb{1 .5 0}{#1}}}
	\par\nobreak\bigskip\medskip\area=0}
\def\chapskip{\par\ifnum\area=0\bigskip\medskip\goodbreak
	\else\newpage\fi}
\def\chapy#1{\area=1\cross=0
	\xdef\lefthead{\rgbo{1 0 .5}{#1}}\vbox{\biggerer\offinterlineskip
	\line{\chrulefill¼\hphantom{\lefthead}\chrulefill}
	\line{\chrulefill¼\lefthead\chrulefill}}\par\nobreak\medskip}
\def\chap#1\par{\chapskip\bookmark3{#1}\chapy{#1}}
\def\sectskip{\par\ifnum\cross=0\bigskip\medskip\goodbreak
	\else\newpage\fi}
\def\secty#1{\cross=1
	\xdef\righthead{\rgbo{1 0 1}{#1}}\vbox{\bigger\offinterlineskip
	\line{\cdotfill¼\hphantom{\righthead}\cdotfill}
	\line{\cdotfill¼\righthead\cdotfill}}\par\nobreak\medskip}
\def\sect#1 #2\par{\sectskip\bookmark{#1}{#2}\secty{#2}}
\def\subsectskip{\par\ifdim\lastskip<\medskipamount
	\bigskip\medskip\goodbreak\else\nobreak\fi}
\def\subsecty#1{\noindent{\sectfont{\rgbo{.5 0 1}{#1}}}\par\nobreak\medskip}
\def\subsect#1\par{\subsectskip\bookmark0{#1}\subsecty{#1}}
\long\def\x#1 #2\par{\hangindent2\parindent%
\mark{0 0 1}\rgboo{0 0 1}{\bf Exercise #1}\\#2%
\par\rgboo{0 0 0}\mark{0 0 0}}
\def\refs{\bigskip\noindent{\bf \rgbo{0 .5 1}{REFERENCES}}\par\nobreak\medskip
	\frenchspacing \parskip=0pt \refrm \baselineskip=1.23em plus 1pt
	\def\ital##1Õ{{\refit##1\/}}}
\long\def\twocolumn#1#2{\hbox to\hsize{\vtop{\hsize=2.9in#1}
	\hfil\vtop{\hsize=2.9in #2}}}


\twelvepoint
\font\bigger=cmbx12 \sca2
\font\biggerer=cmb10 \sca5
\font\biggest=cmssdc10 scaled 4100
 \sca5

 \sca3


\def Ü{\relax\ifmmode\Rightarrow\else\expandafter\subsect\fi}
\def Û{\relax\ifmmode\Leftrightarrow\else\expandafter\sect\fi}
\def Ú{\relax\ifmmode\Leftarrow\else\expandafter\chap\fi}

\def\itemize#1 {\item{\bf#1}}
\def\itemizze#1 {\itemitem{\bf#1}}
\def\itemutem{\par\indent\indent \hangindent3\parindent \textindent}
\def\itemizzze#1 {\itemutem{\bf#1}}
\def ª{\relax\ifmmode\leftrightarrow\else\itemizze\fi}
\def Á{\relax\ifmmode\gets\else\itemizzze\fi}

\def\tbt#1#2#3#4{\left({#1\atop#2}{#3\atop#4}\right)}

\def\¢{\ominus}

\def\Ä{\varphi}  \def\¿{\varpi}	\def\Ï{\vartheta}

\def ò{\relax\ifmmode\cdots\else\dotfill\fi}

\chardef\slo="1C


\def\cvrule{\rgbo{0 .5 1}{\vrule}}
\def\chrule{\rgbo{0 .5 1}{\hrule}}
\def\boxit#1{\leavevmode\thinspace\hbox{\cvrule\vtop{\vbox{\chrule%
	\vskip3pt\kern1pt\hbox{\vphantom{\bf/}\thinspace\thinspace%
	{\bf#1}\thinspace\thinspace}}\kern1pt\vskip3pt\chrule}\cvrule}%
	\thinspace}
\def\Boxit#1{\noindent\vbox{\chrule\hbox{\cvrule\kern3pt\vbox{
	\advance\hsize-7pt\vskip-\parskip\kern3pt\bf#1
	\hbox{\vrule height0pt depth\dp\strutbox width0pt}
	\kern3pt}\kern3pt\cvrule}\chrule}}


\def\boxeq#1{\boxit{${\displaystyle #1 }$}}          


\def\today{\ifcase\month\or
 January\or February\or March\or April\or May\or June\or July\or
 August\or September\or October\or November\or December\fi
 \space\number\day, \number\year}

\parindent=20pt
\newskip\normalparskip	\normalparskip=.7\medskipamount
\parskip=\normalparskip	



\catcode`\|=\active \catcode`\<=\active \catcode`\>=\active 
\def|{\relax\ifmmode\delimiter"026A30C \else$\mathchar"026A$\fi}
\def<{\relax\ifmmode\mathchar"313C \else$\mathchar"313C$\fi}
\def>{\relax\ifmmode\mathchar"313E \else$\mathchar"313E$\fi}


%
%
%
%
%
%
%

\def\thetitle#1#2#3#4#5{
 \def\titlefont{\biggest} \font\footrm=cmr10 \font\footit=cmti10
  \twelverm
	{\hbox to\hsize{#4 \hfill YITP-SB-#3}}\par
	\vskip.8in minus.1in {\center\baselineskip=2.2\normalbaselineskip
 {\titlefont #1}\par}{\center\baselineskip=\normalbaselineskip
 \vskip.5in minus.2in #2
	\vskip1.4in minus1.2in {\twelvebf ABSTRACT}\par}
 \vskip.1in\par
 \narrower\par#5\par\unnarrower\vskip3.5in minus3.3in\eject}
\def\paper\par#1\par#2\par#3\par#4\par#5\par{
	\thetitle{#1}{#2}{#3}{#4}{#5}} 
\def\author#1#2{#1 \vskip.1in {\twelveit #2}\vskip.1in}
\def\YITP{C. N. Yang Institute for Theoretical Physics\\
	State University of New York, Stony Brook, NY 11794-3840}
\def\WS{W. Siegel\footnote{$*$}{
	\pdflink{mailto:siegel@insti.physics.sunysb.edu}\\
	\pdfklink{http://insti.physics.sunysb.edu/\~{}siegel/plan.html}
	{http://insti.physics.sunysb.edu/\noexpand~siegel/plan.html}}}


\pageno=0

\paper

{\rgb{0 0.8 1}{Embedding vs.¼6D twistors}}

\author\WS\YITP

12-11

May 1, 2012

We review the relation between the ``embedding" formalism and spinorial projective space.  The latter is more convenient when treating spin (and indispensable for supersymmetry), as it maintains manifest conformal symmetry while using 4-dimensional indices on fields/operators.  It does this by solving all algebraic constraints using 6-dimensional (off-shell) twistors.  In an added note we review the supersymmetric generalization, and give some new results for N=3.

\pageno=2

ÜProjective lightcone and HP(1)

The projective lightcone [1] (recently dubbed ``embedding formalism" for purposes of vagueness) manifests conformal symmetry SO(D,2) (i.e., makes the coordinates a representation instead of a nonlinear realization) by treating D-dimensional spacetime as a (D+2)-dimensional lightcone, with all points on any ray identified (``projective").  This constraint + gauge invariance eliminates the 2 extra dimensions while preserving the manifest symmetry, until these conditions are solved:  In lightcone notation,
$$ Y^2 = 0âÜâY^A = (Y^+,Y^a,Y^-) = {\bf e}(1,x^a,üx^2) $$
$$ ÜâdY^2 = {\bf e}^2 dx^2,ââYÉY' = -ü{\bf ee}'(x-x')^2 $$
Nonlocal conformal invariants are constructed from inner products of $Y$'s by preserving local scale invariance at each point, thus canceling all factors of the coordinate/local scale factor/worldline einbein {\bf e} (rather than gauging it to 1).  Similarly, covariants of the right weights for scalar operators come from assigning appropriate weights to them, and thus $e$ dependence, through the homogeneity constraint:
$$ \left( YÉ{»\over »Y} + ¶\right) ì = 0âÜâì(Y) = {\bf e}^{-¶} Ä(x) $$

Another projective construction is familiar for SO(3), namely CP(1).  Its Wick rotation RP(1) realizes SO(2,1) on a single real coordinate (e.g., as applied to the boundary of the open string).  Rather than starting with a null 3-vector, it begins with an SL(2,R) spinor, but again with a local scale invariance.  The relationship is clear, since any null 3-vector can be expressed as the ``square" of a 3D spinor (twistor).  An immediate advantage of working with the unconstrained spinor variable is that projection, while yielding the usual nonlnear realization of SO(2,1) (or SO(3) in the complex case), automatically recognizes it as fractional linear transformations:
$$ x' = {ax+b\over cx+d} $$
(In general, projective spaces can also be considered as coset spaces; the result is identical, but requires introducing and then eliminating additional coordinates.  In this case, we would start with a 3-dimensional space with 2 gauge invariances.)

We skip further details of this example to discuss the case of SO(4,2) in D=4.  There the projective space is HP(1), as applied to field theory for constructing general instanton solutions in Yang-Mills [2].  (It can be generalized directly to the supersymmetric case in superspace [3].  There one sees that the generalization of $Y$ is too cumbersome.)  Explicitly, we start with a null SO(4,2) 6-vector, which in SU(2,2) spinor notation is an antisymmetric bi-spinor satisfying
$$ Y^2 ­ \f14 ·_{MNPQ}Y^{MN}Y^{PQ} = 0âÜâY^{MN} = C^{Œº}Â_Œ{}^M Â_º{}^N $$
where $M,...$ are SU(2,2) indices, $Œ,...$ are SL(2,C) indices, $·$ is the usual antisymmetric symbol (which here acts as the 6D Minkowski metric in spinor notation), and $C$ is also an (Hermitian) antisymmetric symbol.  (This actually gives only the half-lightcone, which is OK since the origin needs to be excluded anyway.  The same construction, but with different reality properties, applies to Euclidean space.  Note that in the supersymmetric case there is no analog to $·$, but only graded antisymmetrization of indices, which does not yield just a scalar.  Hence the solution there is only in terms of 6D supertwistors, at least for the case of chiral superspace [3].)

We''ll also need the charge conjugate:  As usual, conjugating $Y$ is the same as ``dualizing" with $·$, so
$$ Y^2 = üY^{MN}ÑY_{MN},ââÑY_{MN} ­ ç_{MÀP}ç_{NÀQ}(Y*)^{ÀPÀQ} = ÐC_{ÀºÀŒ}ÐÂ{}_M{}^{ÀŒ}ÐÂ{}_N{}^{Àº},ââç = \tbt 0{iC}{-iC}0 $$
where $ç$ is the SU(2,2) metric.  The result is that the charge conjugate of $Â$ is not independent, but orthogonal:
$$ Â_Œ{}^M ÐÂ_M{}^{Àº} = 0 $$
As for $Y$, the solution to this constraint reveals $x$:
$$ Â_Œ{}^M = (Â_Œ{}^µ, Â_Œ{}^{Àµ}) = u_Œ{}^à (¶_Ã^µ, x_Ã{}^{Àµ}) ,ââ
	ÐÂ_M{}^{ÀŒ} = (ÐÂ_µ{}^{ÀŒ}, ÐÂ_{Àµ}{}^{ÀŒ}) = (-x_µ{}^{ÀÃ}, ¶_{Àµ}^{ÀÃ}) Ðu_{ÀÃ}{}^{ÀŒ} $$
or in matrix notation
$$  = u ( Iâ-x ),ââР= { x \choose I } Ðu $$
where $u$ is a local GL(2,C) transformation (compensator).  Without loss of generality, we can gauge away the phase piece, so
$$ det(u) = det(Ðu) = {\bf e} $$

The 2$ð$2 matrix $x$ is thus again a ratio, but now of matrices:  Its conformal transformation is as usual nonlinear, but fractional linear:
$$ ÐÂ' = \tbt acbd ÐÂâÜâx' = (ax+b)(cx+d)^{-1} $$
which can also be written as
$$ \tbt acbd^{-1} = çgÿç^{-1} ­ \tbt{÷d}{-÷c}{-÷b}{÷a} $$
$$ ÜâÂ' =  \tbt{÷d}{-÷c}{-÷b}{÷a}âÜâx' = (x÷c+÷d)^{-1}(x÷a+÷b) $$
For the case of spin, it is also useful to have the other transformation laws
$$ Ðu' = (cx+d)Ðu,ââu' = (x÷c+÷d)u $$
In particular, the inversion is the case
$$ \tbt a c b d = \tbt 0 I {-I} 0 $$

ÜSpin

Other than the nice transformation law, this seems like just extra work, until spin is considered.  A useful analogy is general relativity, where vierbeins are somewhat superfluous without spinors.  The analogy goes further for supergravity:  In superspace the coordinates carry superindices, while the fields carry merely tangent-space local-Lorentz indices.  While the vielbein can be used to convert between the two, the tangent-space indices are necessary for considering constraints, actions, etc.

In our case similar remarks apply even without supersymmetry.  (This is also true for coset spaces.)  If spin indices are taken as 6D, then they must be constrained.  Such constraints take the general form [4], in 6D-vector notation,
$$ (S_A{}^B Y_B + ¶'Y_A) ì = 0 $$
for spin operator $S$, where $¶'$ is related to the representation.  ($¶'=¶_0-D/2$, where $¶_0$ is the conformal weight of the corresponding free field.)  They have the same form as general free field equations in D dimensions, but with 4D spin replaced with 6D, momentum $p_a$ replaced with $Y_A$, etc.

For example, a p-form in D=4 becomes a 6D (p+1)-form:  Thus a 4D selfdual 2-form becomes a 6D selfdual 3-form.  Solution of the constraints in reducing from D=6 to 4 is similar to solving free field equations in momentum space, reducing from D=4 to 2 transverse.  So a 6D 3-form Abelian Maxwell field strength would reduce to a 4D 2-form off shell, which in turn would reduce to a 2D 1-form on-shell gauge field.  (The 6D Maxwell field strength in 6D-vector notation is $Y_{[A}»_B A_{C]}=üL_{[AB}A_{C]}$, $YÉA=0$.) 

Rather than giving further details on this lightcone style reduction from D=6 to 4, we instead give the simpler, manifestly conformal twistor solution in the general case:
$$ ì_{M...N}{}^{P...Q}(Y) = ÐÂ_M{}^{ÀŒ}...ÐÂ_N{}^{Àº}Â_©{}^P...Â_¶{}^Q ì_{ÀŒ...Àº}{}^{©...¶} (Â,ÐÂ) $$
where the operators are totally symmetric in all lower indices, and in all upper indices.  (Because of the constraint, any antisymmetric pair of 6D indices can be factorized as a $Y$.)  For example, the 6D 2-form $ì_M{}^N$ reduces to the 4D vector $ì_{ÀŒ}{}^º$, and the selfdual part of the 3-form $ì^{MN}$ reduces to the 4D selfdual part of the 2-form $ì^{Œº}$. In practice, we simply Óstart with operators carrying local 4D indicesÕ.  This manifestly preserves conformal invariance because these indices transform only under the local tangent-space SL(2,C) (and scale) and not under the global conformal SU(2,2).

Since only $u$ and $Ðu$, not $x$, transform under the local tangent-space symmetries (especially 4D Lorentz), the operators with local indices can then be related to 4D operators as
$$ \boxeq{ ì_{ÀŒ...Àº}{}^{©...¶} (u,Ðu,x) = {\bf e}^{-÷¶}Ðu_{ÀŒ}{}^{Àµ}...Ðu_{Àº}{}^{ÀÃ}u_§{}^©...u_ {}^¶ Ä_{Àµ...ÀÃ}{}^{§... }(x) } $$
where $÷¶$ is the weight $¶$ + ($ü$ the number of indices), because $u$ and $Ðu$ implicitly contain a $å{\bf e}$.  We have used the slight trick of replacing $u^{-1}$'s with $u$'s via 
$$ u^{-1} = Cu^T C/det(u) $$
The (nonlinear) conformal transformation of $Ä$ follows from those of $u$ and $Ðu$, and the fact that $ì$ is a conformal scalar.  Thus $u$ and $Ðu$ are like vielbeins that convert ``flat" indices $Œ,ÀŒ$ of the SL(2,C) tangent space to ``curved" indices $µ,Àµ$ of the coordinates $x^{µÀµ}$.  (Similar remarks apply to tangent-space scale symmetry and ${\bf e}=det(u)=det(Ðu)$.)

ÜCorrelators

In the usual approach, correlators for scalars are constructed from inner products $Y_iÉY_j$, replacing the condition of conformal invariance with the simpler one of local scale invariance.  (Equivalently, one could use ${\bf e}_i {\bf e}_j (x_i-x_j)^2$ directly.)  But for correlators with spin, there is still the construction of covariants with various types of indices, carried by $Y$'s, satisfying the constraints (but even that is simpler in spinor notation).  The twistor method is simpler, since there are only 2-component SL(2,C) indices to deal with, rather than 4-component SU(2,2), and no constraints, although there is now the local Lorentz invariance to preserve.

We first note that conformal invariants are all of the form
$$ Â'ÐÂ = u'(x-x')Ðu $$
These are 2$ð$2 matrices.  Their free indices are local, and can be identified with those on the operators.  In particular, we can take the determinant
$$ det(Â'ÐÂ) = det(u')det(x-x')det(Ðu) = ü{\bf ee}'(x-x')^2 = -YÉY' $$

The entire procedure is then to take these invariants 
$$ \boxeq{ u'(x-x')Ðuâ(or¼{\bf ee}'(x-x')^2) } $$
and Ómatch factors of $u$ and $Ðu$ (and their determinant {\bf e}) at each point with those in the operatorsÕ $ì$, as they appear in their relation to $Ä$.  For example, for the general 2-point correlator [5] the result can be seen immediately by inspection:  After peeling off the $u$'s, $Ðu$'s, and $e$'s,
$$ Ò Ä_{Àµ...ÀÃ}{}^{§... }(x)¼Ä_{Àµ'...ÀÃ'}{}^{§'... '}(0) Ô ¾ { x^{(§}{}_{Àµ'}...x^{ )}{}_{ÀÃ'}x^{(§'}{}_{Àµ}...x^{ ')}{}_{ÀÃ}\over (x^2)^{÷¶}} $$
In particular, this can be related to the more cumbersome 4D vector notation with a little algebra:  For example, for the case of the vector,
$$ Ò Ä_{Àµ}{}^µ (x)¼Ä_{ÀÃ}{}^Ã(0) Ô ¾ x^µ{}_{ÀÃ}x^Ã{}_{Àµ} = x^µ{}_{Àµ}x^Ã{}_{ÀÃ} - üC^{µÃ}ÐC_{ÀµÀÃ}x^2 = x^a x^b - üú^{ab}x^2 $$

ÜFurther reading

The generalization to the supersymmetric case (along with its relation to the similar coset approach) was reviewed in [6] (of which this paper is basically a truncation to the nonsupersymmetric case), with references to earlier work, and to the coset approach.  The coset approach was applied to N=4 supersymmetric correlators in [7].  Recently similar results were found using the embedding approach for N=1 [8].

ÜSummary

The easiest way to treat spin for conformal symmetry is to begin with operators carrying local 4D 2-component spinor indices.  There are no constraints to solve, and (as usual) the 2-component spinor algebra is easier than the alternatives.  Generalization to supersymmetry involves mostly just an extension of the range of indices.

ÜAdded note

Finally we give a brief ``review" of the generalization to superspace.  (The real analytic case was reviewed in [6].)  We again construct 6D (super)twistors as rectangular matrices, now with 1 global (P)SU(2,2|N) and 1 local GL(2|n,C) index for the defining representations, where n $²$ N/2.  The 3 most important cases are
$$ \li{ chiral: &ân = 0\cr chiral¼analytic: &ân = {N-1\over 2},ââN¼odd\cr real¼analytic: &ân = {N\over 2},ââN¼even\cr} $$
We start with a (2|n)$ð$(4|N) matrix $Â$ and define its charge conjugate $ÐÂ$ using the U(2,2|N) metric $ç$:
$$ Р­ çÂÿ $$
(Due to the huge number of types of indices in the general case, in this section we stick to matrix notation.)  The solution to the constraint
$$ ÂÐÂ = 0 $$
is then
$$  = u \pmatrix{ I & -÷w & -w +ü÷w÷{Ðw} \cr} ,ââР= \pmatrix{ w +ü÷w÷{Ðw} \cr ÷{Ðw} \cr I \cr} Ðu $$
where the size and content of each submatrix are
$$ \matrix{ submatrix & size & x's & Ï's & ÐÏ's & y's \cr
	u, Ðu & (2|n)^2 & (gauge) &&& \cr
	w & (2|n)^2 & 4 & 2n & 2n & n^2 \cr
	÷w & (2|n)ð(0|N-2n) & 0 & 2(N-2n) & 0 & n(N-2n) \cr
	÷{Ðw} & (0|N-2n)ð(2|n) & 0 & 0 & 2(N-2n) & n(N-2n) \cr} $$
(The $y$'s are R-symmetry coordinates.)  Whereas $u,Ðu$ and $÷w,÷{Ðw}$ are charge conjugate pairs, $w$ is self-conjugate and square.  The above solution is in the real representation; for the chiral and antichiral representations we make the replacements:
$$ \li{ chiral: &âw £ w + ü÷w÷{Ðw} \cr
	antichiral: &âw £ w - ü÷w÷{Ðw} \cr} $$
For the real analytic case, there is no $÷w$:  Operators can be taken to depend on either $Â$ or $ÐÂ$ (and thus only $w$, up to a coordinate gauge transformation).  In the other cases, there are chiral (analytic) operators that depend only on $Â$ ($w$ and $÷w$), and antichiral (analytic) operators that depend on only $ÐÂ$ ($w$ and $÷{Ðw}$), with corresponding restrictions on 4D (local) indices.  But there are also operators that depend on the union of these 2 spaces (i.e., $Â,ÐÂ$, or $w,÷w,÷{Ðw}$), such as the product of chiral (analytic) and antichiral (analytic) operators.  This larger space is the usual full superspace (without $y$'s) for the chiral (not analytic) case.  For the special case of N=3 super Yang-Mills, the field strength lives on the chiral analytic space, while the prepotentials live on the larger superspace [9].  

Conformal invariants are all constructed from the $(2|n)^2$ matrices
$$ Â'ÐÂ = u'( w -w' -÷w'÷{Ðw} +ü÷w÷{Ðw} +ü÷w'÷{Ðw}' )Ðu $$
Thus there is translation invariance in all the 4D coordinates in the real analytic case.  In the other cases, we can drop the 2 latter terms if the latter (primed) coordinates are put in the chiral representation, and the former (unprimed) in the antichiral (which is useful only if the corresponding operators live on the corresponding chiral/antichiral superspaces).  As before, a scalar invariant is the superdeterminant of this expression.  It gives the free propagators for scalar field strengths for N=0,1,2,3,4 in appropriate superspaces.  As usual, for other scalar operators we can take appropriate powers of it, restricted by scale weight (i.e., canceling {\bf e}'s), to find general multi-point correlators.  Except for N=0 or 4, there is a U(1) in the local group that restricts correlators for operators that live in the larger (chiral + antichiral) spaces.

A particularly simple case is chiral (not analytic) operators, which are known to carry only undotted spinor indices and no R-symmetry indices.  This ties in directly with the supertwistor construction, since $Â$ (chiral) carries only undotted SL(2,C) indices, while $ÐÂ$ (antichiral) carries only dotted.  Thus the general chiral-antichiral 2-point correlator is again obvious:
$$ ÒÄ^{µ...Ã}(x,Ï)¼ÐÄ_{À§...À }(x',ÐÏ')Ô ¾ {öx{}^{(µ}{}_{À§}...öx{}^{Ã)}{}_{À }\over (öx{}^2)^{÷¶}} $$
where $öx$ is the 2$ð$2 matrix given above ($x-x'+ÏÐÏ'$ in the chiral representation for $Ä$, antichiral for $ÐÄ$).  Similar remarks apply to chiral analytic operators, but instead of SL(2,C) indices they carry SL(2|${N-1\over 2}$,C) indices:  For example, for N=3, these are SL(2|1,C) indices, consisting of an undotted spinor index together with a single-valued R-symmetry index (and the charge conjugate for antichiral analytic).  These replace the indices in the 2-point correlator above, while $öx$ itself is replaced with $w-w'+÷w÷{Ðw}'$ in the (anti)chiral analytic representation.

ÜAcknowledgment

This work is supported in part by National Science Foundation Grant No.¼PHY-0969739.

\refs

£1 P.A.M. Dirac, ÓAnn. Math.Õ É37 (1936) 429;\\
H.A. Kastrup, ÓPhys. Rev.Õ É150 (1966) 1183;\\
G. Mack and A. Salam, ÓAnn. Phys.Õ É53 (1969) 174;\\
S. Adler, ÓPhys. Rev.Õ ÉD6 (1972) 3445;\\
R. Marnelius and B. Nilsson, ÓPhys. Rev.Õ ÉD22 (1980) 830.

£2 M.F. Atiyah and R.S. Ward, ÓComm. Math. Phys.Õ É55 (1977) 117;\\
M.F. Atiyah, V.G. Drinfel'd, N.J. Hitchin, and Yu.I. Manin, ÓPhys. Lett.Õ É65A (1978) 185;\\
E. Corrigan, D. Fairlie, P. Goddard, and S. Templeton, ÓNucl. Phys.Õ ÉB140 (1978) 31;\\
N.H. Christ, E.J. Weinberg, and N.K. Stanton, ÓPhys. Rev.Õ ÉD18 (1978) 2013;\\
M.F. Atiyah, ÓGeometry of Yang-Mills fieldsÕ (Scuola Normale Superiore, Pisa, 1979);\\
V.E. Korepin and S.L. Shatashvili, ÓMath. USSR IzvestiyaÕ É24 (1985) 307.

£3 
  W. Siegel,
  ÓPhys. Rev.Õ ÉD47 (1993) 2512
  \xxxlink{hep-th/9210008};
  ÓPhys. Rev.Õ  ÉD52 (1995) 1042
  \xxxlink{hep-th/9412011}.

£4 W. Siegel, ÓIntroduction to string field theoryÕ (World Scientific, 1988),\\
\xxxlink{hep-th/0107094}, sect. 2.2.

£5 
  D. Simmons-Duffin,
 \xxxlink{1204.3894} [hep-th].

£6 
  W. Siegel,
  \xxxlink{1005.2317} [hep-th].

£7 
  P.S. Howe and P.C. West,
  ÓInt. J. Mod. Phys.Õ ÉA14 (1999) 2659
  \xxxlink{hep-th/9509140};
  ÓPhys. Lett.Õ ÉB400 (1997) 307
  \xxxlink{hep-th/9611075};\\
  B. Eden, P.S. Howe, and P.C. West,
  ÓPhys. Lett.Õ ÉB463 (1999) 19
  \xxxlink{hep-th/9905085}.

£8 
  W.D. Goldberger, W. Skiba, and M. Son,
  \xxxlink{1112.0325} [hep-th].

£9 
  A. Galperin, E. Ivanov, S. Kalitsyn, V. Ogievetsky, and E. Sokatchev,
  ÓClass. Quant. Grav.Õ  É2 (1985) 155;\\
  A.S. Galperin, E.A. Ivanov, and V.I. Ogievetsky,
  ÓSov. J. Nucl. Phys.Õ  É46 (1987) 543
   [ÓYad. Fiz.Õ  É46 (1987) 948];\\
  F. Delduc and J. McCabe,
  ÓClass. Quant. Grav.Õ  É6 (1989) 233.

\bye